\documentclass[prl, showpacs, twocolumn, superscriptaddress]{revtex4}
\usepackage{graphicx}

\begin{document}

\title{Intermittent dynamics and $1/f^\beta$ noise in single cardiac muscle cells}

\author{Tomomi Yokogawa}
\author{Takahiro Harada}
\email[Corresponding Author: ]{harada@life.ne.his.fukui-u.ac.jp}
\affiliation{Department of Human and Artificial Intelligent Systems, University of Fukui, Fukui 910-8507, Japan}

\date{\today}

\begin{abstract}
Fluctuations in the spontaneous beating activity of isolated cardiac cells were studied over a timescale of six decades.
The beat dynamics of single cardiac cells were heterogeneous and intermittent.
The interbeat intervals (IBIs) were power-law distributed in a long-time regime.
Furthermore, for long timescales up to the experimental window, the autocorrelation of IBIs exhibits a scaling behavior of $1/f^\beta$-noise type.
These observations suggest that $1/f^\beta$ noise is an intrinsic characteristic of spontaneous activity of single cardiac cells.
\end{abstract}


\maketitle

Recently, it is rapidly recognized that noise is ubiquitous in cellular biology.
Owing to the development of single-molecule experiments on various kinds of proteins, their stochastic motion has been extensively studied \cite{Ritort:2006}.
It has also been observed that several cellular behaviors, including gene expression \cite{Elowitz:2002} and motile activity \cite{Cluzel:2004}, exhibit large noise.
The properties and the significance of such cellular noise are currently of great interest in both physics and biology.

It is also well known that a variety of signals from multicellular organisms exhibit characteristic fluctuations \cite{Glass:2000}.
A typical example of such fluctuations is that observed in human heart-beat rate \cite{Musha:1982, Peng:1993}.
It is now widely accepted that the power spectrum of the human heart beat rate exhibits $1/f^\beta$ dependence ($\beta \simeq 1$) in the low-frequency regime.
However, in spite of intensive studies in the last two decades \cite{Goldberger:2002}, the mechanism of $1/f^\beta$ noise generation still remains unclear.

In the preset Letter, in order to reveal the properties of the cellular noise in cardiac cells and to understand their relevance to the fluctuations at the multicellular level, we study the fluctuations in the activity of cardiac muscle cells at the single-cell level.
Although there have been several studies reporting that the tissue culture of the cardiac muscle cells also exhibit scale-invariant fluctuations \cite{Kucera:2000, Soen:2000, Yoneyama:2004, Pzharskii:1998}, it is unknown whether $1/f^\beta$ noise is an intrinsic property of a single isolated cardiac cell.

In order to clarify this point, we studied the spontaneous activity of single isolated cardiac muscle cells in a wide range of time scales, ranging from subseconds to several days.
It is then found that the spontaneous beating activity of single isolated cells exhibits strong heterogeneity and intermittency.
The probability distribution of the interbeat interval (IBI) exhibits a power-law dependence in the long-time regime.
Furthermore, a long-range correlation analysis, termed detrended fluctuation analysis (DFA) \cite{Peng:1994}, indicates that the time series of IBI fluctuations exhibit scale invariance over a timescales of three decades and are characterized as $1/f^\beta$ noise.
This experimental finding suggests that the $1/f^\beta$ noise is characteristic of an isolated single cardiac cell.

\begin{figure}[b]
\begin{center}
\includegraphics[scale=1.0]{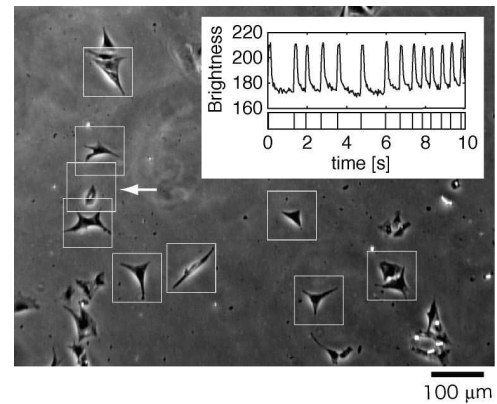}
\caption{Phase-contrast image of isolated cells in culture. The cells whose spontaneous activity was studied are marked by solid squares. The scale bar represents 100 $\mu$m. Inset: A part of the time series of 8-bit brightness for the cell indicated by the white arrow (top). Raster plot constructed from the time series of brightness (bottom). The timings of the spontaneous contractions are marked by vertical lines.}
\label{f.illust}
\end{center}
\end{figure}
\begin{figure*}[t]
\begin{center}
\includegraphics[scale=1.0]{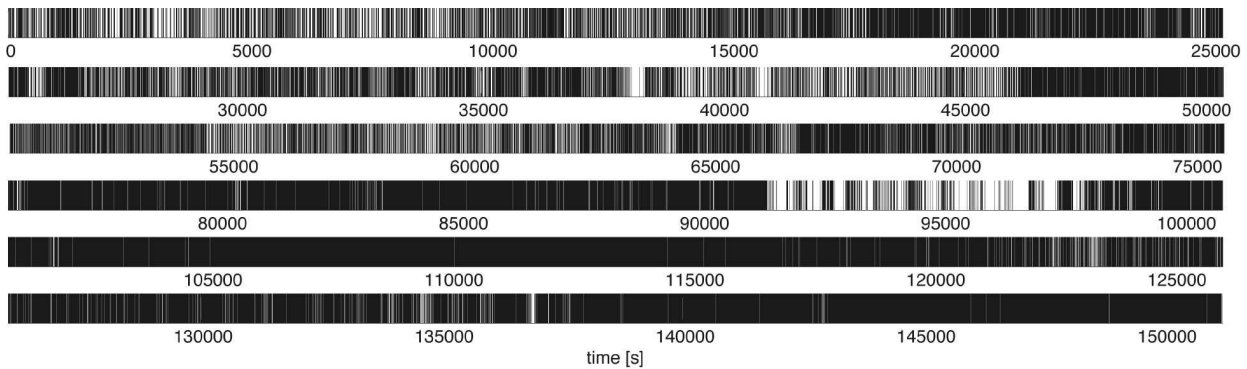}
\caption{Typical raster plot of a single cell. Each vertical line represents the timing of a single contraction. The numbers associated with this plot indicate the time elapsed since the onset of measurement.}
\label{f.whole}
\end{center}
\end{figure*}
Spontaneous beating activity was examined for single isolated ventricular cells derived from neonatal rats.
The cells were plated on a petri dish, which is coated with collagen, at a low density (30 cells/mm$^2$) \cite{preparation}.
At this density, most of the cells have no contact with other cells (see Fig.~\ref{f.illust}). The cardiac muscle cells, which were identified by their morphology and ability to display spontaneous contraction, were isolated from each other during the experiment.
The culture dish was set in a customized incubation chamber placed on an inverted microscope, and the cells were cultivated at a constant condition (37$^\circ$C, 5\% CO$_2$, saturated humidity).
Phase-contrast images of cells were obtained via a charge-coupled device camera at 30 frame/s and were stored on a hard disk for the following analysis.

Contractions of a cardiac muscle cell cause a temporal change in the brightness of the phase-contrast images.
Thus, the timings of the spontaneous contractions of the cell were determined from the time series of the brightness of a pixel in the cell \cite{spikes}.
The inset in Fig.~\ref{f.illust} displays a typical example of a time series of brightness of a single cell, presented with a raster plot (timings of contractions) determined from the original time series.
By employing the above mentioned experimental setup, the profiles of spontaneous activities of 11 cells from three different experiments have been analyzed.
For each cell, data were obtained for several hours (mean 32.5 h, ranging from 6 to 55 h).


Figure \ref{f.whole} displays a typical example of the raster plot of a single cell.
As observed in this plot, the timings of contractions are heterogeneous.
We found that there are several typical patterns in this trace.

\begin{figure}[b]
\begin{center}
\includegraphics[scale=1.0]{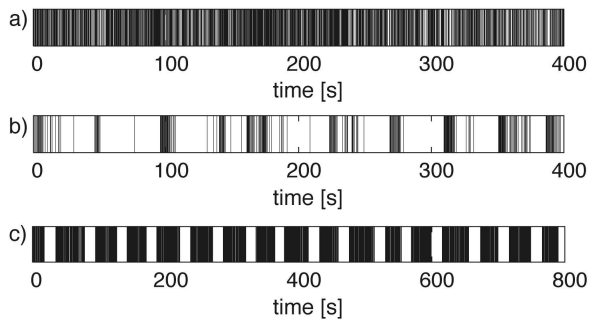}
\caption{Magnified raster plots of several typical temporal patterns; (a) Pattern A shows steady contractions with rather small fluctuations and (b) pattern B depicts an intermittent bursting pattern. The durations of the active and quiescent phases vary stochastically. (c) Pattern C shows a regular bursting pattern. Active and quiescent phases occurs alternately with a regular periodicity.}
\label{f.zoom}
\end{center}
\end{figure}

First, steady contractions with some fluctuations were frequently observed, as observed in the portion around 110000 s in Fig.~\ref{f.whole}. This pattern is represented in Fig.~\ref{f.zoom}a and is termed pattern A.
It is also found that contractions often become intermittent. Bursts of contractions are separated by quiescent periods.
This pattern is observed around 40000 s in Fig.~\ref{f.whole}, for instance, and is shown at a higher resolution in Fig.~\ref{f.zoom}b. This pattern is termed B.
In this pattern, the duration of quiescent periods occasionally reaches several hours.
We also notice that in some cases, the bursting activity becomes quite regular. This pattern is represented in Fig.~\ref{f.zoom}c.
In this case, quiescent periods occur with high regularity, although the periods of alternation vary slowly. This pattern is termed C.

Although we can categorize the patterns of the contraction as mentioned above, this categorization seems to be rather subjective.
In the entire time series for the contractions, a pattern is gradually replaced with another pattern, and subsequent patterns do not have distinct boundaries.
It might be more appropriate to define the patterns in terms of the duration of the quiescent periods: small or no quiescent periods (pattern A), quiescent periods with durations ranging from several seconds to several minutes (pattern B), regular quiescent periods (pattern C).

\begin{figure}[tb]
\begin{center}
\includegraphics[scale=1.0]{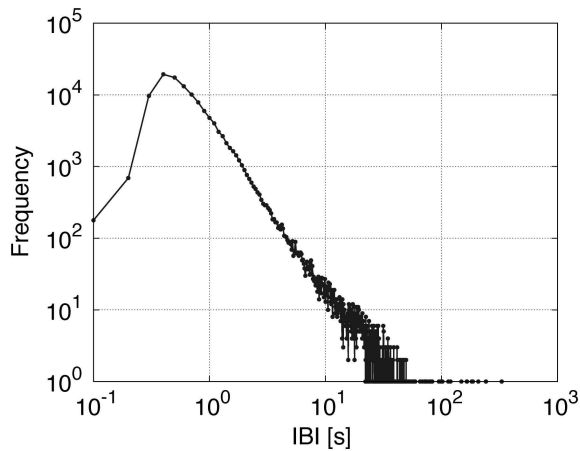}
\caption{Frequency of IBIs in a logarithmic plot. A fit in the region $[10^{-0.5}, 10^{1.5}]$ yields the exponent $-2.23 \pm 0.02$.}
\label{f.hist}
\end{center}
\end{figure}

In order to characterize the statistics of the IBIs, we calculated their distribution.
Figure \ref{f.hist} displays the distribution of IBIs for the same cell as that in Fig.~\ref{f.whole} on the logarithmic scale.
It is found that a peak appears around 0.5 s, which indicates the presence of a refractory period shorter than 0.5 s.
Furthermore, for larger timescales, a scaling behavior is observed over a timescale of two decades.

The exponent of scaling seems to depend on the patterns of activity of each cell.
For cells that exhibit a large fraction of pattern B (intermittent), the exponent is close to 2.
For cells exhibiting a large fraction of pattern A (regular), the scaling behavior becomes less distinct and the exponent increases.
Moreover, for cells that exhibit a large fraction of pattern C (regular burst), the second peak appears around several tens of seconds, which is interpreted as the modal length of the quiescent periods.
The mean value of the exponent averaged over the all cells was $-3.1 \pm 1.2$.

Next, in order to study the two-point statistics of the IBI time series, we performed a long-term memory analysis (the DFA) \cite{Peng:1994}.
This method enables us to investigate the autocorrelation of a time series by removing the effect of nonstationarity.
First, letting $\tau_i$ represent the $i$-th IBI, the IBI time series are integrated as $x_i \equiv \sum_{j = 1}^i \tau_j$.
Subsequently, the entire time series is divided into $m$ subsections of size $n$; in each of these subsections, the linear local trend is calculated by means of the least-square fit.
After the local trend is subtracted in each subsection, the variance of the residual time series is calculated.
Therefore, the quantity to be analyzed here is represented as
\begin{equation}
V(n) \equiv \frac{1}{m n} \sum_{j = 1}^{m} \sum_{i = j n + 1}^{(j+1)n} \left( x_i - \bar x^j_i \right)^2,
\label{e.dfa}
\end{equation}
where $\bar x^j_i$ represents the local trend in the $j$-th subsection.
If the time series has scale invariance, this quantity scales with $n$ as
\begin{equation}
V(n) \propto n^\alpha.
\end{equation}
The exponent $\alpha$ obtained in this manner is related to the scaling exponent $\beta$ of the power spectrum of the time series $x_i$ as $\alpha = \beta + 1$.

\begin{figure}[tb]
\begin{center}
\includegraphics[scale=1.0]{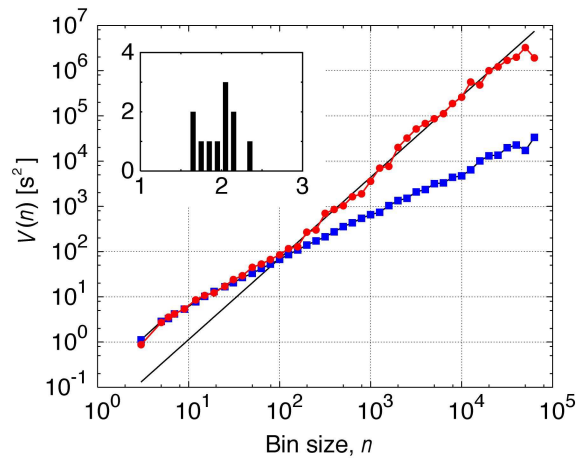}
\caption{(Color online) Typical result of the DFA for a single cell. Solid circles represent $V(n)$ calculated for the original data obtained in the experiment. The solid line is a fit in the region $[10^{2}, 10^{4.5}]$, and its slope yields the scaling exponent $\alpha =1.79 \pm 0.03$.
Solid squares represent $V(n)$ calculated for the surrogate (randomly shuffled) data, and its slope yields the exponent close to unity ($\alpha = 0.96 \pm 0.02$).
Inset represents the frequency of the exponents in the long-term scaling region.}
\label{f.dfa}
\end{center}
\end{figure}

Figure \ref{f.dfa} displays a typical example of $V(n)$, which is calculated for the same cell as that chosen to plot Fig.~\ref{f.hist}, plotted against $n$ on the logarithmic scale.
Two scaling regions are prominent in this plot.
One is the short-time region ($n \le 10^2$), where the exponent is $\alpha \simeq 1$.
The other is the long-time region ($n \ge 10^3$), where the exponent is $\alpha \simeq 2$.
Therefore, the fluctuations in the IBI time series are uncorrelated for short time scales (shorter than $10^2$ beats).
For time scales larger than $10^3$ beats, it has been found that the IBI time series has a power-law correlation, and this correlation can be characterized as $1/f^\beta$ noise.
This power-law correlation seems to be sustained until at least $10^5$ beats, although this limit is set by the period of the experiment (55 h).

For comparison, we prepared a randomly shuffled IBI time series as surrogate data.
As expected, the surrogate data yields $V(n)$ with a single scaling exponent, close to unity for all the time scales.
From Fig.~\ref{f.dfa}, it is apparent that the slope of the original data is distinct from that of the surrogate data.

In order to examine the cell-to-cell variance of the abovementioned tendency, the frequency of $\alpha$ is presented in the inset of Fig.~\ref{f.dfa}.
The value of the exponent is determined by fitting $\log V(n)$ to $\alpha \log n + b$ in the scaling region $n > 10^3$. 
From this figure, it is found that the value of $\alpha$ is around 2.
The mean value of $\alpha$ was $1.97 \pm 0.23$.
This suggests that the $1/f^\beta$-noise-type long-term correlation is a generic property observed in the activity of single cardiac muscle cells.

The abovementioned experimental findings suggest that intermittent, power-law distributed, and power-law correlated IBIs are the inherent properties of a single cardiac muscle cell.
Here, let us summarize earlier studies concerning the fluctuations of IBIs in a cardiac tissue. 
Kucera {\it et al.} reported that the beat rates exhibit a $1/f^\beta$ spectrum in a dense network of cardiac muscle cells, although it was not clear in their report whether it is an intrinsic property of single cardiac muscle cells or a phenomenon that arises through interactions among cells \cite{Kucera:2000}.
Yoneyama {\it et al.} performed the DFA for isolated cardiac cells and reported that the exponent $\alpha/2$ is close to 0.5 in the short-time regime ($n < 10^2$), which is consistent with our result \cite{Yoneyama:2004}.

Soen {\it et al.} studied the beatrate fluctuations in single isolated cardiac cells in terms of the Allan Factor (AF) measure \cite{Soen:2000}.
In their report, they reported that the IBIs are uncorrelated in the time scale $t < 10^2 \times$ (mean IBI), which is consistent with our result; however, IBIs exhibit strong anti-correlation for larger time scales. This observation corresponds to $\alpha \simeq 3$ for the DFA for a long time scale, which differs from our result.
One possible reason for this discrepancy might be the difference in the quantity analyzed. The essential difference between the DFA and the AF measure is that the former deals with the beat number as a measure of time, while the latter deals directly with the physical time.
When applied to a highly heterogeneous time series, the DFA is preferred to AF measure.
The difference between both measures is evident when applied to the surrogate data. As observed in Fig.~\ref{f.dfa}, the DFA yields nice scaling with $\alpha \simeq 1$ for the all time scales, while the AF provides significantly less clear scaling for the surrogate data (see Ref.~\cite{Soen:2000}).

In conclusion, it has been found that a single cardiac muscle cell exhibits heterogeneous and intermittent dynamics.
The IBIs obey a power-law distribution, and the time series of the IBIs are power-law correlated and exhibit $1/f^\beta$ noise in the long-time regime.
This correlation is sustained for at least several tens of hours.
These properties of single cardiac cells might affect the heartbeat dynamics {\it in vivo}.

The mechanism behind these singular properties at the single-cell level is an open problem.
The fact that $1/f^\beta$ noise is stably observed implies the presence of a robust mechanism for these intermittent fluctuations.
With regard to this problem, it has been reported that membrane currents in several types of cells exhibit a power-law correlation \cite{Teich:1997}.
In the investigation of the mechanism of the power-law correlation in single-cell fluctuations, cardiac muscle cells have a great advantage because one can investigate the IBI time series from their contractile motion in a noninvasive manner, and this fact facilitates parallel and extremely long-term measurements.
This advantage will enable us to clarify the fundamental mechanism behind power-law correlations in the cellular activity at the single-cell level.

\begin{acknowledgments}
The authors are grateful to acknowledge valuable supports from K.~Murase and H.~Ikeda at the University of Fukui and K.~Yoshikawa at Kyoto University.
This work was partly supported by grants from the Ministry of Education, Science, Sports, and Culture, Japan (No.~19031010) and the Research and Education Program for Life Science at the University of Fukui.
\end{acknowledgments}


\begin{thebibliography}{99}
\bibitem{spikes} The timings of contractions were determined as follows.
First, for each cell to be analyzed, a pixel that underwent brightness modulation with a sufficiently large amplitude was automatically selected for every 20 minutes, and the temporal difference of brightness was calculated for the pixel.
In the time-differentiated time series, peaks that exceed a certain threshold were assigned a single contraction.
The value of the threshold was determined manually for each cell and was updated every 20 minutes so that the detected timings of contraction coincide with those inferred visually.
\end{thebibliography}
\end{document}